\documentclass[7pt,a4paper,twocolumn]{article}
\usepackage[utf8]{inputenc}
\usepackage[T1]{fontenc}
\usepackage{mathptmx}
\usepackage{graphicx}
\usepackage[colorlinks=false, pdfborder={0 0 0}]{hyperref}
\usepackage{calc}
\usepackage{enumitem}

\usepackage[a4paper, lmargin=0.08\paperwidth, rmargin=0.08\paperwidth, tmargin=0.08\paperheight, bmargin=0.08\paperheight]{geometry}

\usepackage[all]{nowidow}
\usepackage[protrusion=true,expansion=true]{microtype}
\usepackage{caption}

\usepackage{amsmath}
\usepackage{amsthm}

\usepackage{amsfonts}
\usepackage{siunitx}
\usepackage{complexity}
\usepackage{nicefrac}
\usepackage{subcaption}
\usepackage{svg}

\usepackage{titlesec}
\titlespacing*{\section}{0pt}{1pt}{1pt}
\titlespacing*{\subsection}{0pt}{1pt}{1pt}
\titlespacing*{\paragraph}{0pt}{1pt}{1pt}

% ...
\usepackage{booktabs}
\usepackage{multirow}
\usepackage{numprint}
\npdecimalsign{.}

\usepackage{niceAlgorithm}
\SetKwFunction{PrunedBFS}{PrunedBFS}
\SetKwFunction{Preprocess}{Preprocess}
\SetKwFunction{TopoSweep}{TopoSweep}
\SetKwFunction{prune}{prune}
\SetKwFunction{continue}{continue}
\SetKwFunction{seen}{seen}
\usepackage{float}

\newcommand{\vertices}{V}
\newcommand{\edges}{E}

\newcommand{\hlLabel}{L}
\newcommand{\bitset}{b}
\newcommand{\fwdbitset}{\bitset_{{\mathrm{fwd}}}}
\newcommand{\bwdbitset}{\bitset_{{\mathrm{bwd}}}}
\newcommand{\fwdLabel}{\hlLabel_{{\mathrm{fwd}}}}
\newcommand{\bwdLabel}{\hlLabel_{{\mathrm{bwd}}}}

\author{Steil, Patrick}
\title{Parallel Hub Labeling for Reachability on DAGs}
\makeatletter

\usepackage{fancyhdr}
\begin{document}
    \section*{Introduction.}

    Here we present a parallel variant of PLL for reachability on DAGs.
    This variant was developed by the author during a seminar at the Karlsruhe Institute of Technology (KIT), with the specific goal of reducing the preprocessing for hub labeling on large time-expanded graphs; which model public transit networks.

    \subsection*{The Original PLL Algorithm.}
    Pruned Landmark Labeling (PLL)~\cite{DBLP:conf/sigmod/AkibaIY13} is a  sequential HL algorithm that computes a minimal labeling given a vertex ordering (see Algorithm~\ref{alg:ppl:preprocess}).

    \begin{algorithm}[h]
        \caption{Preprocess - taken from~\cite{DBLP:conf/sigmod/AkibaIY13}}
        \label{alg:ppl:preprocess}
        \myproc{$\Preprocess(G)$}{
            $\hlLabel_{0}[v] \gets \emptyset\;\forall v \in \vertices$\;
            \ForEachComment{by importance}{$k \in [1, n]$}{
                $\hlLabel_{k} \gets \PrunedBFS(G, v_k, \hlLabel_{k-1    })$\;
            }
            \Return $\hlLabel_{n}$\;
        }
    \end{algorithm}
    
    An important subroutine is a modified BFS (or \textit{for the weighted case} Dijkstra), which prunes the search space using already computed labels (see Algorithm~\ref{alg:ppl:prunedbfs}).

    \begin{algorithm}[h]
        \caption{PrunedBFS - taken from~\cite{DBLP:conf/sigmod/AkibaIY13}}
        \label{alg:ppl:prunedbfs}
        \myproc{$\PrunedBFS(G, v_k \in \vertices, \hlLabel_{k-1})$}{
            $Q \gets \left\{v_k\right\}$\;
            $\seen \gets \left\{v_k\right\}$\;
            $\hlLabel_{k}[v] \gets \hlLabel_{k-1}[v]\;\forall v \in \vertices$\;
            \While{$Q \neq \emptyset$}{
                Dequeue $u$ from $Q$\;
                \IfComment{prune}{$\hlLabel_{k-1}[v_k] \cap \hlLabel_{k-1}[u] \neq \emptyset$}{\continue}
                $\hlLabel_{k}[u] \gets \hlLabel_{k}[u] \cup \{v_k\}$\;
                \ForEach{$(u, v) \in \edges \;\mathrm{s.t.}\; v \notin \seen$}{
                    $Q \gets Q \cup \left\{v\right\}$\;
                    $\seen \gets \seen \cup \left\{v\right\}$\;
                }
            }
            \Return $\hlLabel_{k}$\;
        }
    \end{algorithm}

    \subsection*{Parallelization Challenges.}
        
    As the algorithm is inherently sequential, parallelization is not trivial.
    Note that for directed graphs, HL stores two labels per vertex, $\fwdLabel$ and $\bwdLabel$.
    Algorithmically, this means the BFS searches have to performed twice; once on the forward graph $G$, and once on the reverse graph $G^{\mathrm{rev}}$.
    The algorithm is inherently sequential, and simply parallelizing the BFS (see Algorithm 1) only provides a limited speedup, since the average node degree is relatively low.
    Running $\PrunedBFS$ from multiple vertices simultaneously loses the pruning property, and can produce very large labels under certain circumstances.
		
    Instead of parallelizing a single BFS, multiple BFSs are executed simultaneously.
    To allow for pruning, a thread $t$ processing $\PrunedBFS(v_t)$ must determine during the relaxation of an edge $(u, v) \in \edges$ whether the path $v_t \to v$ is already covered or if another thread will cover it (using a more important hub).
    To achieve this, we introduce the following approach:
    
    \subsection*{New Parallelization Scheme.}
    We use two bitsets $\bwdbitset, \fwdbitset: \vertices \to 2^{K}$.
    After Algorithm~\ref{alg:ppl:toposweep}, bitset $\bwdbitset[u][j]$ stores whether the $j$th starting vertex can reach $u$, and analogously $\fwdbitset[u][j]$ whether $u$ can reach the $j$th starting vertex.
    When thread $t \in [1, K]$ processes vertex $v_t \in \vertices$, we extend the pruning rule of Algorithm~\ref{alg:ppl:prunedbfs} by the following condition when relaxing edge $(u, v) \in \edges$:
		
    If $\exists t' \in [1, t)$, such that $\fwdbitset[v_k][t']$ and $\bwdbitset[v][t']$, then $v_k \to v$ will be covered by $t'$, as $v_t \to v_{t'} \to v$.
    This can be easily checked by using bit-operations.
    
    We note that Algorithm~\ref{alg:ppl:toposweep} is highly cache efficient, and the operations on bitsets have a high throughput.
    
    The batch size $K$ and thread assignment strategies can be tuned experimentally; we set $K \in \{128, 256, 512\}$, in order to make use of registers.
    For larger graphs, the $\mathcal{O}(m)$ topological sweep may become inefficient later on in the main algorithm.
    We therefore relax the pruning and process the unimportant vertices in parallel.
    While this may no longer yield strictly minimal labels, the increase in redundant labels is very small, as most of the hubs are created during the first few important searches.
			
    The code for this project can be found here\footnote{\url{https://github.com/PatrickSteil/DAG-HL}}.
    
    \begin{algorithm}[h]
        \caption{Topological Sweep}
        \label{alg:ppl:toposweep}
        \myproc{$\TopoSweep(G, b, \left<v_1, \dots, v_K\right>)$}{
            $\bwdbitset[v], \fwdbitset[v] \gets 0,\;\forall v \in \vertices$\;
            \ForEach{$i \in [1, K]$}{
                $\bwdbitset[v_i][i], \fwdbitset[v_i][i] \gets 1$\;
            }
            \ForEach{$(u, v) \in \edges$\,sorted topologically}{
                $\bwdbitset[v] \gets \bwdbitset[v] \mathbin{|} \bwdbitset[u]$\;
            }
            \ForEach{$(u, v) \in \edges$\,sorted topologically in rev}{
                $\fwdbitset[u] \gets \fwdbitset[v] \mathbin{|} \fwdbitset[u]$\;
            }
            \Return $\bwdbitset, \fwdbitset$\;
        }
    \end{algorithm}
    \bibliographystyle{plainurl}
    \bibliography{bibliography.bib}
\end{document}